\documentclass[fleqn,12pt,twoside]{article}
\usepackage{espcrc1}
\usepackage{graphicx}
\title{
\hfill{\small {\bf MKPH-T-03-14}}\\[0.5cm]
\bf Final state interaction effects in $\eta$ photoproduction on two-
  and three-body nuclei}
\author{A. Fix\address[MCSD]{Institut f\"ur Kernphysik, 
        Johannes Gutenberg-Universit\"at Mainz, \\ 
        D-55099 Mainz, Germany} and 
        H. Arenh\"ovel\addressmark}
\begin{document}
\maketitle

\begin{abstract}
The role of final state interaction in $\eta$ photoproduction on a deuteron as
well as on three-body nuclei is investigated within few-body scattering
theory. Deviation of the
theory from the available experimental results is briefly discussed.   
\end{abstract}

\section{Incoherent $\eta$ photoproduction on a deuteron}
Photoproduction of $\eta$ mesons on the deuteron near threshold is strongly
influenced by final state interaction (FSI). 
The main reason for the importance of FSI is a strong mismatch
between a large momentum transfer and a minimal kinetic energy in the
final $\eta NN$ system, which on the other hand can effectively be balanced
by the interaction between the final particles. 
Furthermore, an appreciable attraction in the $\eta NN$ system  
generates a virtual pole in the three-body scattering 
amplitude \cite{FiAr00}, which in turn leads to a strong rise of
the cross section just above threshold in agreement with experimental
results \cite{Kru95}. 

The calculation was performed 
using a separable representation of the driving $\eta N$ and $NN$ scattering
amplitudes as described in detail in \cite{FiAr00}. 
For the $\eta N$ interaction an isobar ansatz of \cite{BeTa91} was used with
inclusion of $S_{11}(1535)$ only. The corresponding parameters were adjusted
such that the $\eta N$ scattering length $a_{\eta N}=(0.5+i0.32)$ fm,
which we consider as an approximate average of the scattering lengths 
provided by modern $\eta N$ analyses, is reproduced 
and that at the same time a good description is provided 
of the processes $\pi N\to\pi N$, $\pi^-p\to \eta n$, and $\pi N\to\pi\pi N$. 
The $NN$-interaction was considered only in the dominant $s$ wave scattering
states. We have used the version BEST3 for $^1S_0$ and
correspondingly BEST4 for the triplet states.
The separable representation allows one to 
reduce the three-body problem to a set of integral equation in one
variable only, whose structure is analogous to the usual
Lippmann-Schwinger equation for two coupled channels $(\eta+d)$ and $(N+N^*)$.

\begin{figure}[htb]
\begin{center}
\includegraphics[width=15.5pc]{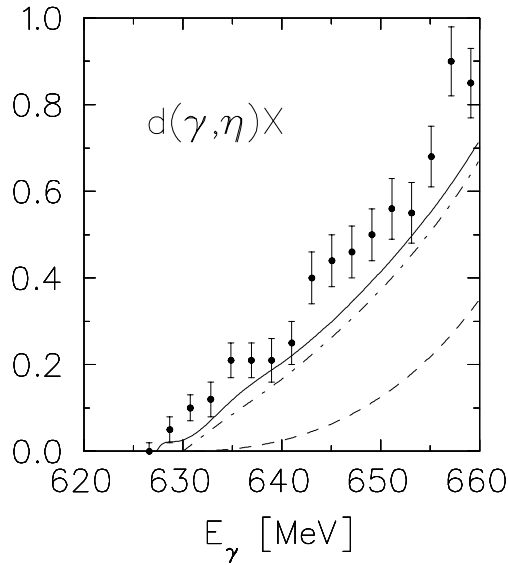}
\vspace{-0.9cm}
\end{center}
\caption{Total cross section for $\gamma d\to\eta X$ (solid line). The
  dash-dotted (dashed) curve is obtained for the reaction $\gamma d\to\eta np$
  with (without) FSI. The data are from \protect\cite{Kru95}.}
\label{fig1}
\end{figure}

The resulting inclusive cross section for $\gamma d\to \eta X$
($X = \{np,d\}$) is shown by the solid curve in Fig.\,\ref{fig1}. As one can
see, although the three-body calculation leads to a sizeable
improvement as compared to the impulse approximation (dashed
line), a quantitative agreement with the data is not yet achieved. 
A possible reason for this disagreement could be the neglect of $\pi NN$ 
configurations in the calculation. 
The contribution of the intermediate pions depends strongly on the 
role of large momentum transfers in the reaction, since it is associated with
a large intermediate momentum and consequently is effective only at short
distances. For example, inclusion of 
$\pi NN$ configurations provides only a small fraction of 
the $\eta d$ scattering cross section \cite{FiAr00}. At the same time 
they contribute rather sizeably to 
coherent $\eta$ photoproduction on the deuteron \cite{RiAr00}, where the large
transferred  
momenta emphasize naturally the short distances in the
target. In view of the systematic deviation between theory and the data, the
role of intermediate pions in the incoherent channel $\gamma d\to \eta np$
must be considered in greater detail. 

\section{Coherent $\eta$ photoproduction on $^3$He}

\begin{figure}[htb]
\begin{minipage}[t]{75mm}
\includegraphics[width=15.5pc]{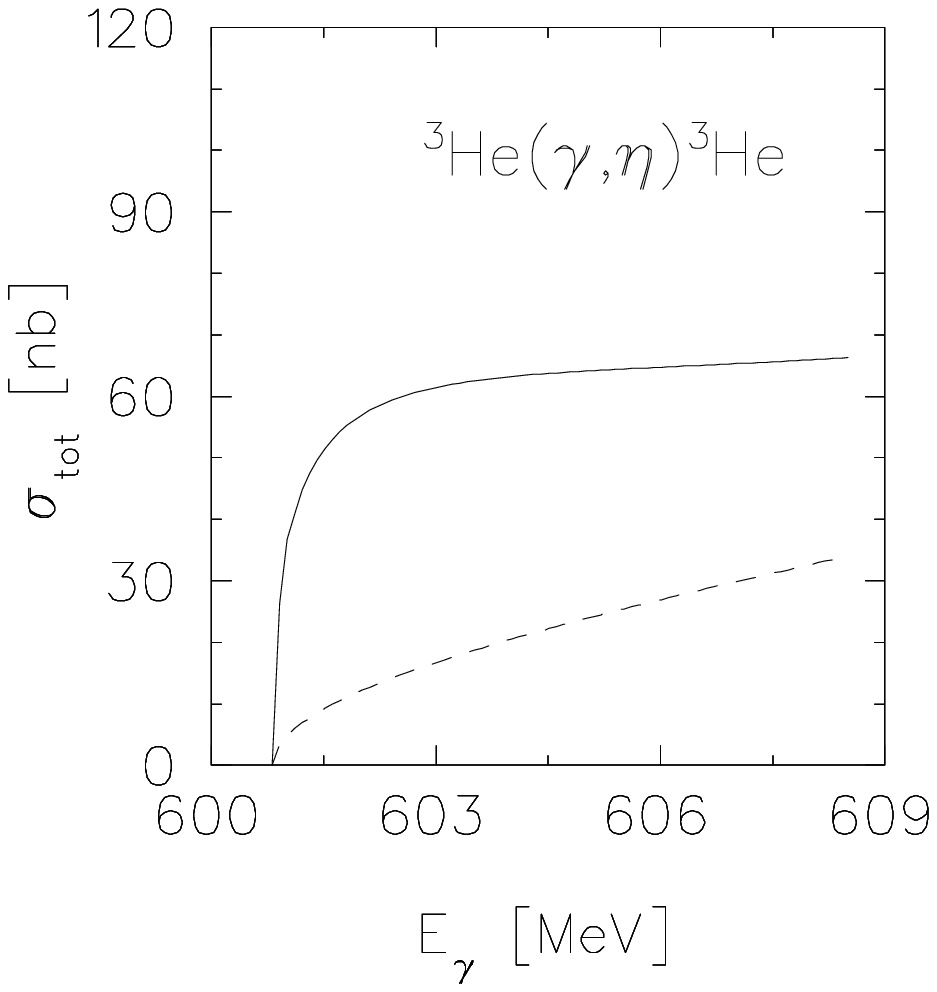}
\vspace{-0.9cm}
\caption{Total cross section for the photoproduction of $\eta$ on
  $^3$He. Solid (dashed) curve presents the result with (without) FSI.} 
\label{fig2} 
\end{minipage}
\hspace{\fill}
\begin{minipage}[t]{75mm}
\includegraphics[width=15.5pc]{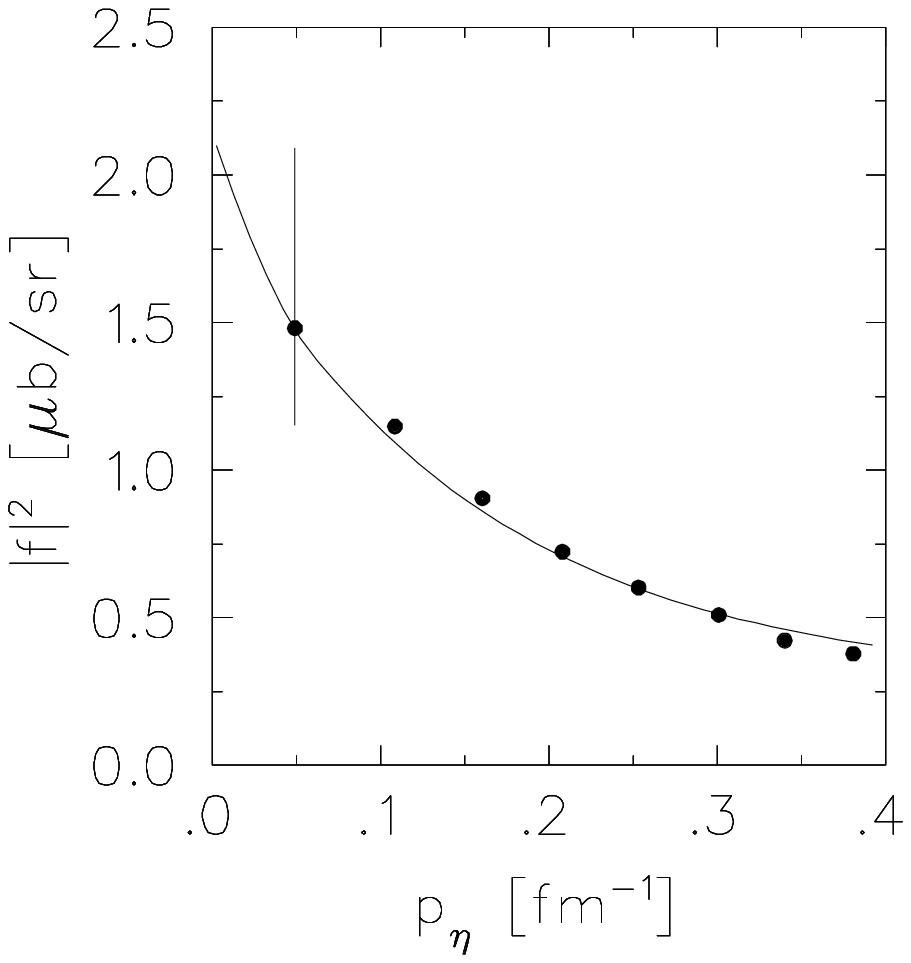}
\vspace{-0.9cm}
\caption{Production amplitude squared for the reaction 
$^3$He$(\gamma,\eta)^3$He compared with the experimental results for
  $d(p,\eta)^3$He  \protect\cite{May96}. The theoretical result is arbitrarily
  normalized to the  data at $p=0.25$ fm$^{-1}$.}  
\label{fig3}
\end{minipage}
\end{figure}

The study of $\eta\,^3$He elastic scattering, based on four-body
Faddeev-Yakubovsky theory \cite{FiAr02}, shows that there is a 
virtual state in this system, which in turn strongly influences 
low-energy $\eta$ production on three-body nuclei. This feature is
demonstrated in Fig.\,\ref{fig2}, where one can see a very rapid rise of the
total cross section $^3$He$(\gamma,\eta)^3$He, in contrast to a much flatter
form predicted by the plane wave approximation (dashed line). 

Comparing our results with the not yet published TAPS data \cite{Pfeif02} we
find a strong discrepancy. It is rather surprising that 
even the form of the experimental cross section is not
reproduced. According to the Migdal-Watson theory \cite{Mig77} the energy
dependence of the cross section very close to threshold is determined  
primarily by the low-energy parameters of $\eta\,^3$He scattering
and is therefore more or less universal for different entrance channels. For
instance, as is demonstrated in Fig.\,\ref{fig3}, the energy dependence of the
squared amplitude calculated for the reaction $^3$He$(\gamma,\eta)^3$He
section agrees quite well with the one extracted from the data of
$d(p,\eta)^3$He \cite{May96}. From this
viewpoint the TAPS data could be explained only by assuming another long-range
mechanism of $\eta$ production which does not contribute to the 
reaction $d(p,\eta)^3$He.

In summary, the three-body aspects of FSI are very
important for $\eta$ production on light nuclei near threshold. 
The results of an exact
three-body treatment exhibit a systematic deviation from the data. We suspect
that a possible reason of this disagreement might be a pion rescattering
mechanism, whose role up to now has not been investigated in detail. In the
coherent reaction on 
$^3$He the FSI effect is also of fundamental importance. The
energy dependence of the cross section, which is determined by the low-energy
parameters of $\eta\,^3$He elastic scattering agrees reasonably well with that
observed in the $pd$ collision. On the other hand the very strong energy
dependence of the experimental cross section for $^3$He$(\gamma,\eta)^3$He is
not explained. We think, that this disagreement requires further
investigations on the theoretical as well as on the experimental side.


\begin{thebibliography}{9} 
\bibitem{FiAr00} A. Fix and H. Arenh\"ovel, Nucl. Phys. A {\bf 697} (2002)
  277.
\bibitem{Kru95} B. Krusche {\it et al}., Phys. Lett. B {\bf 358} (1995) 40. 
\bibitem{BeTa91} C. Bennhold and H. Tanabe, Nucl. Phys. A {\bf 530} (1991) 62. 
\bibitem{RiAr00} F. Ritz and H. Arenh\"ovel, Phys. Rev. C {\bf 64} (2001)
  034005. 
\bibitem{FiAr02} A. Fix and H. Arenh\"ovel, Phys. Rev. C {\bf 66} (2002)
  024002. 
\bibitem{Pfeif02} M. Pfeiffer, PhD Thesis (Giessen 2002).
\bibitem{Mig77} A.B. Migdal, Front. Phys. {\bf 48} No.\,4 (1977) 1-437. 
\bibitem{May96} B. Mayer {\it et al}., Phys. Rev. C {\bf 53} (1996) 2068.
\end{thebibliography}
\end{document}